\documentclass[%
 aip,
 amsmath,amssymb,
 reprint,%
]{revtex4-2}

\usepackage{graphicx}
\usepackage{dcolumn}
\usepackage{bm}

\usepackage[utf8]{inputenc}
\usepackage[T1]{fontenc}
\usepackage{mathptmx}
\usepackage{etoolbox}
\usepackage{amsfonts}
\usepackage{mathtools}

\DeclarePairedDelimiter\ket{\lvert}{\rangle}
\DeclarePairedDelimiterX\braket[2]{\langle}{\rangle}{#1 \delimsize\vert #2}

\makeatletter
\def\@email#1#2{%
 \endgroup
 \patchcmd{\titleblock@produce}
  {\frontmatter@RRAPformat}
  {\frontmatter@RRAPformat{\produce@RRAP{*#1\href{mailto:#2}{#2}}}\frontmatter@RRAPformat}
  {}{}
}%
\makeatother
\begin{document}

\title[]{Bias, length, or coupling? What controls the quantum efficiency of electroluminescent single-polymers}
\author{Facundo Tarasi}
\affiliation{Departamento de Qu\'{i}mica Inorg\'{a}nica, Anal\'{i}tica
y Qu\'{i}mica F\'{i}sica/INQUIMAE, Facultad de Ciencias Exactas
y Naturales, Universidad de Buenos Aires, Ciudad Universitaria,
Pab. II, Buenos Aires (C1428EHA) Argentina}

\author{Esteban D. Gadea}
\affiliation{Department of Chemistry, The University
of Utah, Salt Lake City, Utah 84112-0850, United States}

\author{Tchavdar Todorov}
\affiliation{Centre for Light-Matter Interactions, School of Mathematics and Physics, Queen’s University Belfast, Belfast BT7 1NN,
United Kingdom}
\author{Dami\'an A. Scherlis}
\email{damian@qi.fcen.uba.ar}
\affiliation{Departamento de Qu\'{i}mica Inorg\'{a}nica, Anal\'{i}tica
y Qu\'{i}mica F\'{i}sica/INQUIMAE, Facultad de Ciencias Exactas
y Naturales, Universidad de Buenos Aires, Ciudad Universitaria,
Pab. II, Buenos Aires (C1428EHA) Argentina}
\affiliation{Condensed Matter and Statistical Physics,
International Centre for Theoretical Physics
I-34151 Trieste, Italy.}

\date{\today}

\begin{abstract}

Since the first evidence of luminescence of organic polymers in STM junctions,
efforts have been invested in elucidating how to leverage the
voltage, anchoring chemistry, and molecular structure to optimize emission power and efficiency.
Understanding the fundamentals underlying current-driven molecular emission
is important not only for OLED engineering, but also to
control luminescence at the atomic scale toward the mastering of single or localized photon sources.
However, the difficulty in isolating the separate roles of the variables at play
in molecular junction experiments, has precluded a general comprehension
of their distinctive effects on the emitted power and the quantum yield.
In the present report, we use time-dependent electronic structure simulations based
on quantum electrodynamics
to disentangle the incidence of bias, electronic coupling and molecular length on device performance,
with polyphenylene-vinylene as a case study.
A careful validation demonstrates that our approach can achieve quantitative
agreement with available experimental data. Through its application
we identify the applied bias as the main factor determining
emission power. The quantum efficiency, however, is influenced only minimally
by bias and electronic coupling, and is instead dominated by polymer length, on which it depends
exponentially. Thus, using longer polymer chains emerges as the primary strategy for achieving higher efficiencies.
Our results thereby provide key prescriptions for designing single-molecule electroluminescent platforms.

\end{abstract}

\maketitle

\section{Introduction}

Light emission induced by a circulating current is termed electroluminescence.
Modern illumination and display technologies rely on this principle,
harnessed by semiconducting devices known as light emitting diodes (LEDs).
Whereas in incandescent lamps the light is also the result of a current,
incandescence is a consequence of the heating of the metallic filament.
Instead, the luminescence of semiconducting  materials 
can be  explained as the spontaneous emission
from the conduction band, populated as a result of the applied voltage.
Thus, the  wavelength of the radiation is determined by the band gap of the semiconductor,
or, more precisely, by its emission spectrum.
Equivalently, the operation mechanism is often
pictured in terms of the holes and electrons 
injected at each end of the device when a voltage is applied, which
recombination results in photon emission \cite{bookelectrl1}.

Electroluminescence was first reported more than a century ago,
for a silicon carbide crystal \cite{carborundum}. Thereafter, a large variety of inorganic materials
were employed to manufacture light emitting diodes of different wavelengths
from the UV to the infrared, but with a focus on the visible part of the spectrum.
LEDs based on doped GaN, GaAs, InGaN or Y$_3$Al$_5$O$_{12}$, are examples
that became  ubiquitous in lighting, screens and digital panels \cite{bookelectrl1}.

In the 80's it was discovered that electroluminescence could also be found
in conjugated organic polymers. Devices of this class were called
OLEDs (organic light emitting diodes), and their discovery immediately generated
significant and growing interest \cite{science_1996, cr_ostroverkhova, oledbook}.
Their versatility that includes tunable emission, processability, mechanical flexibility
and transparency, added to their excelling color quality, potential efficiency and low environmental impact, 
explain why they became so attractive. Their performance, however, still prevents a large-scale replacement
of one technology by the newer one.
In this context, a major goal
is the optimization of the quantum efficiency, defined as the ratio between photons emitted and
charge carriers injected \cite{oledbook}.

A vast body of experimental and theoretical information exists around organic 
electroluminescent molecules \cite{cr_ostroverkhova, oledbook, chemsci_dimaiolo, jacs_147_2239}.
However, along with this knowledge, some fundamental questions that are central to
OLEDs design still persist. Among such questions, we focus here on the following:
how does the luminescent yield depend on the molecular length, 
the applied bias, and the electrical coupling at the boundaries?
These aspects that are critical to OLED performance and engineering are not easily addressed
experimentally, because of the difficulty in varying these parameters in
a controlled way without affecting other variables.
Electroluminescence has been measured for single molecules in STM setups \cite{prl_112_047403, natcomm_15_1677, prr_5_033027,nanolett_20_7600, science_361_6399, Klaus},
but even in these cases a systematic exploration of the impact of length
and coupling strength on the quantum efficiency remains unfeasible.
On the other hand, this phenomenon is also challenging theoretically.
Whereas different approaches have been proposed to tackle light-matter interaction
dynamics in molecular systems \cite{pnas_114_3026, pra_97_032105, jcp_150_044102, prl, chemrev_rubio}, 
the present problem demands the treatment of the radiative
emission in the presence of a current for a realistic $\pi$-conjugated model. 
This is instrumented here through the application of  quantum electrodynamics 
in combination with a self-consistent mean-field tight-binding Hamiltonian 
and electronic open boundaries \cite{jcp_fluor, jcp_electrl}.
With this methodology
we simulate photon emission of a molecule between two metallic contacts subjected to a potential,
elucidating the distinct roles of length, voltage, and interfacial coupling
on luminescence power  and efficiency.

\begin{figure}
    \centering
    \includegraphics{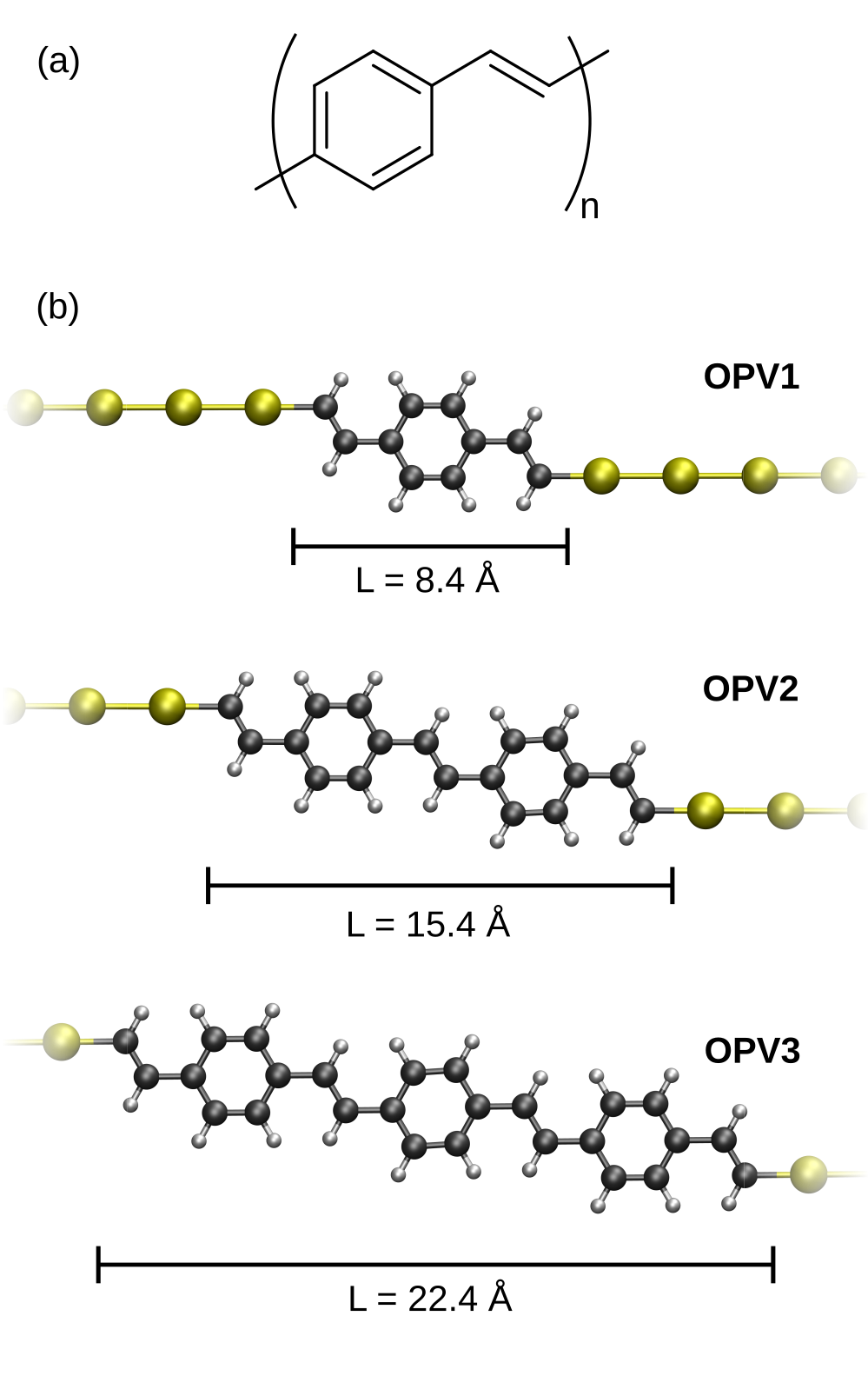}
    \caption{ (a) Chemical structure of the p-phenylenevinylene repeating unit. (b) Ball-and-stick models for the first three oligomers ($n=1, 2, 3$) of the oligo(p-phenylenevinylene) (OPV) series, shown forming a molecular junction between two electrodes (yellow spheres). $L$ is the overall length of the molecular wire.}
    \label{OPV}
\end{figure}

Oligo(p-phenylenevinylene) or OPV (Figure \ref{OPV}) is an electroluminescent organic semiconductor
which was among the first polymers to find application in OLEDs \cite{nature_347_539,acr_1999, oledbook}.
We have chosen it as a case study in our simulations because, while providing
a realistic model of an electroluminescent material with a broad experimental characterization,
its structure allows for a reasonably accurate description
through a relatively simple Hamiltonian. 
Our approach reproduces experimental quantum efficiencies and some of the
fundamental features of OLED behavior. In particular, it tells us that the yield increases
exponentially with the length of the chain, whereas the coupling strength and the applied
bias have less direct and more complex effects.

\section*{Methods}

Simulations of the electron dynamics coupled to a photon bath were carried out in the framework of
the Redfield formalism recently devised by our group \cite{jcp_electrl, jcp_fluor, jcp_qc}.
A self-consistent  tight-binding Hamiltonian was adopted to describe the electronic structure
of oligophenylenvinylene,
with the hopping  terms fitted to reproduce the experimental polymer band gap of 2.4 eV.
The driven Liouville von Neumann approach was employed to induce a current between
two metallic, one-dimensional electrodes of 75 atoms each.
The combination of these strategies has proven capable of reproducing constitutive features in molecular conductance, including Landauer transport, band bending and negative differencial
resistance \cite{jcp_124_214708, jcp_electrl}. Details on the equation of motion and all parameters
involved can be found in Section 1 of the Supporting Information.

\section*{Results and discussion}
\label{results_discussion}

\subsection*{Transport}

\begin{figure}
\centering
\includegraphics{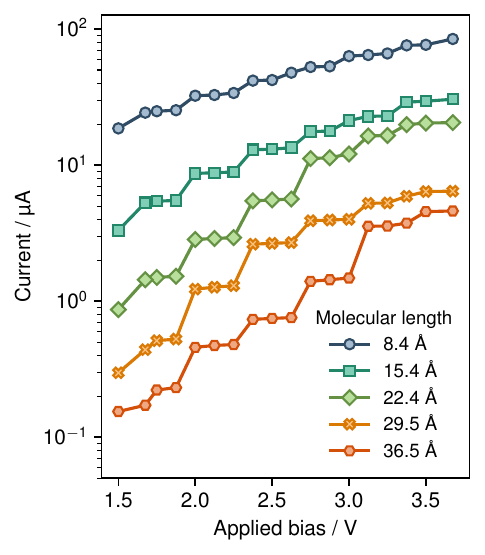}
	\caption{Current as a function of applied bias, for single oligophenylenvinylene molecules with varying lengths as indicated in the legend. }
\label{CvsV}
\end{figure}

Figure \ref{CvsV} shows the current ($I$)
as a function of applied bias $\Delta V$ for  a set of OPV molecules of varying lengths. The series of jumps in the
current is an expected result as new conducting states become 
accessible with the rise in applied voltage. 
Simulated values are consistent with the available
experimental reports on single molecular junctions or
nanometer sized films of OPV or nearly related compounds \cite{nlett_8_1, 
nlett_6_2184, science_320_1482, jacs_124_10654},
and also with previous calculations based on Green's functions \cite{jacs_124_10654, jmc_19_3899, chemphyschem_9_1416}.
Currents typically measured in single strands of three monomers ($\sim$ 23 \AA)
with potentials in the range 0.5 - 1.5 V are in the order of 1 $\mu$A, consistent with
our results \cite{jacs_124_10654, nlett_8_1, nlett_6_2184, science_320_1482}.

\begin{figure}
\centering
\includegraphics{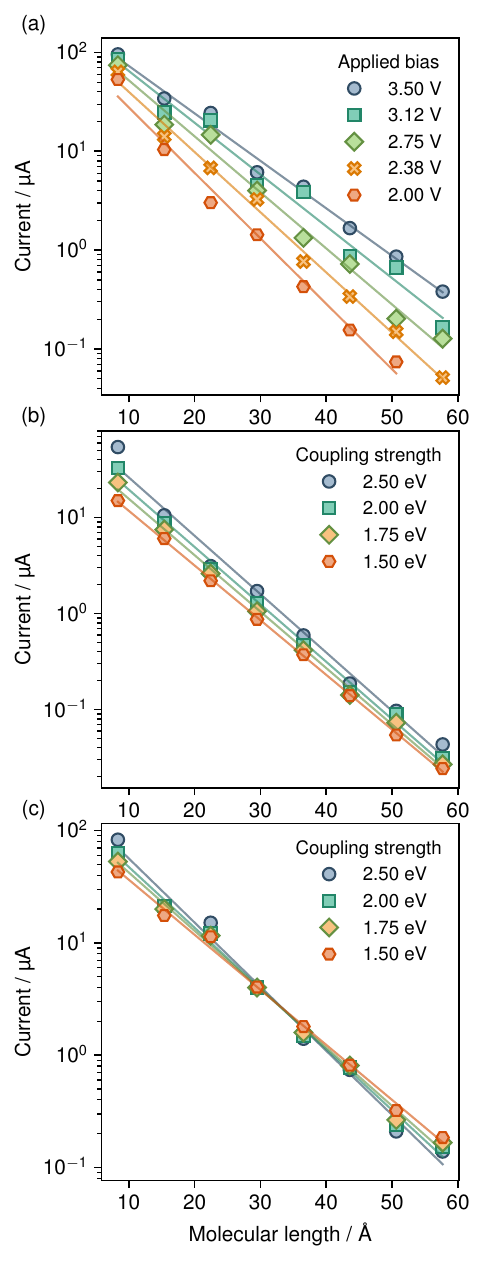}
	\caption{Current as a function of oligomer length. (a) Effect of applied bias, with
	a metal-molecule coupling strength of 2 eV.
	(b) Effect of coupling strength, for an applied bias of 2 V.
	(c) Effect of coupling strength, for an applied bias of 3 V.}
\label{CvsL}
\end{figure}

Figure \ref{CvsL} shows that the current drops exponentially with the length ($L$), in line with a tunneling
transport mechanism. Experiments have established that in long conjugated polymers, of above 3 or 4 nm
depending on temperature,
tunneling is superseded by hopping \cite{jacs_126_4052, science_320_1482, jacs_132_11658, acsnano_3_3861}.
Since we are not including phonons, a hopping mechanism cannot
be captured by our model, and tunneling governs the behavior across the whole polymer series.
The  decay constant ($\beta$) takes values
between 0.1 and 0.2 \AA$^{-1}$ depending on bias and coupling strength.
In particular, $\beta$ turns out to  be quite sensitive to the applied
potential (Figure \ref{CvsL}a), behavior that can be explained by  the increased contribution from higher-energy states within the bias window and by the tilting of the tunneling barrier.
This  decreasing trend  with $\Delta V$ (Section 2 of the Supporting Information) has been observed for various compounds in single molecule
conductance experiments, with conjugated systems exhibiting a stronger variation in comparison
to aliphatic chains \cite{science_301_1221, naturenanot_7_713, jacs_140_12877, prl_112_047403}.
The tunneling decay constants found in our simulations fall close to, though
slightly below, those reported experimentally for OPV and derivatives, which are in the range
0.3 to 0.4 \AA$^{-1}$ \cite{science_320_1482, jacs_128_11260,nlett_5_783}.
In comparing these values,  it must be noted that the experimental data were collected
at low bias, usually below 500 mV, whereas the results from our simulations correspond to
potentials above 1.5 V.
A theoretical estimate for the same polymer combining DFT and Green's function, on the other hand, 
yielded 0.17 \AA$^{-1}$, in  agreement with our calculations \cite{chemphyschem_9_1416}.

The  bond strength at the contacts, which determines the electronic coupling, is another key factor
known to modulate the current \cite{pccp_16_23529}.
This  can be modeled in our simulations through the electrode-molecule coupling term, $\Gamma$.
The range of values explored for this parameter, between 1.5 and 2.5 eV, mimics from weakly to strongly interacting anchoring groups, such as for example carboxylic acid and thiols, exhibiting ionic and covalent binding respectively (see Supporting Information).

In principle, weaker couplings are associated with lower currents, but this may
not be the case in every situation.
Figures \ref{CvsL}b and \ref{CvsL}c illustrate this for $\Delta V$ = 2 V and 3 V.
Our simulations systematically show that $\beta$ becomes smaller with weaker coupling.
The same dependence  has been seen in both experiments and DFT simulations, 
showing an increase of $\beta$ when raising the stability of the anchoring 
group \cite{jacs_128_15874, jacs_135_12228, chemeuj_20_4653, pccp_16_23529, jpcc_113_20967}.
In fact, within limits, the result of higher couplings can be
interpreted as the combination of two effects:  a drop in contact resistance,
and the energetic broadening of the molecular conducting states. The former affects the prefactor of
the exponential decay term, and the latter the exponent itself \cite{pccp_16_23529, jacs_128_15874}.
This behavior is addressed in more detail in Section 2 of the Supporting Information.
As a consequence of this rise in the tunneling decay constant, there should be a length
above which a weaker coupling yields a higher current. This length depends on the applied bias:
in our model, it turns out to be around 3.5 nm for $\Delta V$ = 3 V (Figure \ref{CvsL}c).
In room temperature experiments, presumably, this inversion might be difficult to observe, because it occurs
for lengths at which hopping takes over the conductance.

The overall consistency between our results and the available data---including semi-quantitative
agreement---is  reassuring about the reliability of our approach and 
lends support to the results discussed in what follows.

\subsection*{Emission power}

\begin{figure}
\centering
\includegraphics{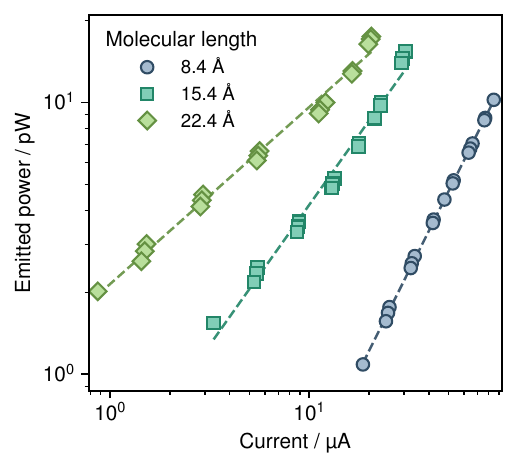}
	\caption{Logarithm of the emitted power versus the logarithm of the current, for a metal-electrode
	coupling of 2 eV, and different oligomer lengths. These plots prove the empirical
	relation $P \propto I^n$ with
	$n$ equal to 1.5, 1.0 and 0.65 for oligomers containing 1, 2 and 3 units.}
\label{PvsI}
\end{figure}

The power radiated in the steady state ($P$) tends to increase with the  applied voltage (see Figure S4, Supporting Information). 
In particular, 
various studies have sought to characterize  $P$ as a function of $I$ in thin films and
single molecules,
identifying a power law relating these two variables, with an exponent in the range  0.6 - 2
\cite{synmet_69_415, natcomm_15_1677, prr_5_033027, angew_136_e202318143}.
This relation between the emitted power and the circulating current is manifest also in our simulations,
with exponents falling in a similar range, depending on polymer length and coupling strength.
Figure \ref{PvsI} presents these results for a molecule-electrode
coupling parameter of 2 eV, and polymers containing 1, 2 and 3 monomers, with fitted exponents equal to 1.5, 1
and 0.65, respectively. These values become smaller for weaker couplings (see Section 4 of the Supporting Information).

\begin{figure}
\centering
\includegraphics{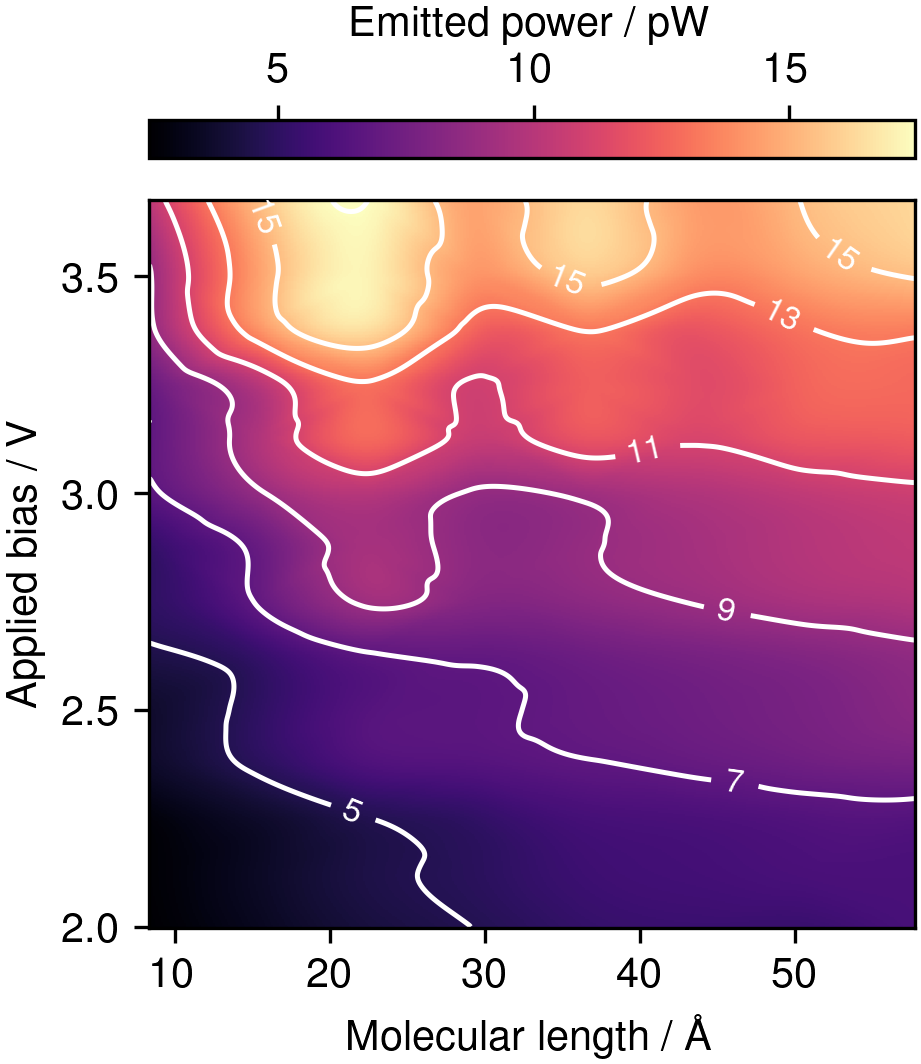}
        \caption{Color map of the power emitted by the electroluminescence of oligophenylenvinylene, 
	 as a function of the applied bias
	and polymer length. The corresponding metal-molecule coupling parameter is 2 eV.}
\label{powermap}
\end{figure}

The behavior of the power as a function of polymer length is not obvious, as it is the outcome of two
opposing influences: on the one hand a longer chain, by
increasing the number of emitters or the cross section,  favors radiative dissipation, while on the other
it promotes the exponential decay of the current. 
The overall effect is thus not easy to predict, and in fact, according to our simulations,
it does not follow a universal trend but turns out to be dependent on the particular
system and conditions \cite{jcp_electrl}. 
Figure \ref{powermap} shows that in the present case the length of the molecule in general enhances
emission, but at high bias the power achieves maximum values for intermediate sizes.
A detailed analysis reveals that these maxima above  2.5 V originate from localized electronic states
of the molecule with high oscillator strengths, which remain unpopulated at lower bias.
These states are not equally present in longer polymers, and thus the effect of the number of
atoms on the emitted power will depend on the specific molecular electronic structure 
(see Section 3 of the Supporting Information for  details).
In any case, the effect of the length on the dissipated electromagnetic radiation
proves to be small in comparison to its effect on the current. This will have a decisive impact
on the quantum efficiencies, as analyzed in what follows.

\subsection*{Quantum efficiency}

A central motivation for the present study is to investigate the quantum efficiency ($QE$)
of electroluminescence and the conditions that maximize it.
Do longer polymers, larger bias, and stronger couplings contribute to the efficiencies?

\begin{figure}
\centering
\includegraphics{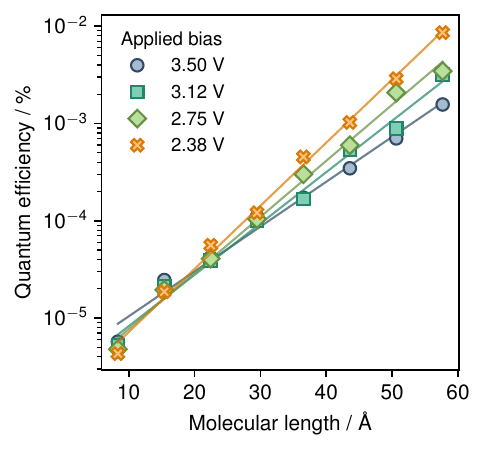}
        \caption{Quantum efficiency of electroluminescence  as a function of oligomer length
	for different applied potentials, and a metal-molecule coupling of 2 eV.}
\label{QvsL}
\end{figure}

As discussed, the emitted power is only slightly sensitive to polymer length,
while the current drops exponentially.
As a result, the  ratio between radiated photons and injected electrons is dominated
by the variation of the latter, 
benefiting from the increase in polymer length due to the drop of the denominator.
Figure \ref{QvsL} displays a quasi exponential behavior of
the efficiency with respect to length, at constant bias.
The slopes of the curves in Figure \ref{QvsL} are essentially the same as those obtained for the tunneling decay
constants from the plot $\log I$ versus $L$ (Figure \ref{CvsL}a) at the same applied bias.

The  efficiencies achieved span several orders of magnitude, from 10$^{-6}$\% to 10$^{-2}$\%,
depending strongly on length and, less markedly, 
on bias and coupling. Single molecule electroluminescence efficiencies
have rarely been quantified in experiments, and in particular we were unable to find
any published  data for OPV in molecular junctions.  
For other molecules, including polythiophene, porphyrine and phtalocyanine,
quantum yields in the range 10$^{-7}$\% -- 10$^{-2}$\% have been reported in 
STM measurements \cite{nanolett_20_7600, prl_112_047403, Klaus}.
Larger values, of around 0.1\%, have been published for macrocycles or chromophores decoupled
from the metallic substrate through different strategies \cite{natcomm_14_8253, natcomm_8_580}.
In these cases the rise in  $QE$ is attributed to the suppression of
the molecular fluorescence quenching, which typically arises from the close contact between the chromophore
and the metal surface \cite{prl_130_036201}. Beyond this, 
we have shown that the smaller currents contribute to maximizing $QE$.

\begin{figure}
\centering
\includegraphics{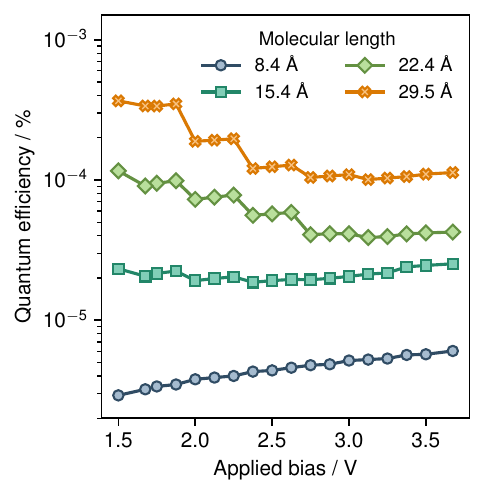}
        \caption{Quantum efficiency of electroluminescence  as a function of applied bias
	for differet oligomers, and a metal-molecule coupling of 2 eV.}
\label{QvsV}
\end{figure}

Now, what is the impact of the applied bias on the electroluminescence efficiency?
This is not a simple question, because $\Delta V$  enhances both the emission and the circulating current, 
and the net effect on their ratio is not evident a priori.
Interestingly, Figure \ref{QvsV} indicates that the answer depends on polymer length. 
Short molecules exhibit higher yields with increasing voltages, but the opposite trend
prevails as the molecule becomes longer. In particular, for a chain
of  3 monomers or more, efficiency improves with lower bias.  

This peculiar response to $\Delta V$  is rooted in the power-law relating $P$ and $I$.
If $P \propto I^n$, it turns out that $QE \propto I^{n-1}$.
Hence, since we are in a regime of positive differential conductance, $I$ always increases with the bias and
if the exponent $n$ is greater than 1, a rise in the applied potential enhances the  efficiency. Conversely, an exponent smaller than 1 has the
opposite effect. When $n$=1, any change in the potential induces a proportional variation in both power
and current, leaving their ratio unchanged. This occurs for $L\approx$15 \AA;
the efficiency in polymers of this length is insensitive to the applied potential.

This exponent is not only a function of the length, but also of the molecule-electrode coupling strength
(see Section 4 of the Supporting Information). As a consequence, the electronic coupling 
determines how the bias affects the quantum efficiency.
At lower couplings, $n$ decreases, meaning that the regime where the bias
favors the efficiency gets narrower.
For example, for a coupling parameter of 1.5 eV, $n$ is equal to or smaller than 1 for all
lengths (Figure S5, Supporting Information), and the electroluminescent yield is optimized at low bias.
For strongly coupled junctions, higher potentials will improve efficiencies, except in the
case of long polymers.

\begin{figure}
\centering
\includegraphics{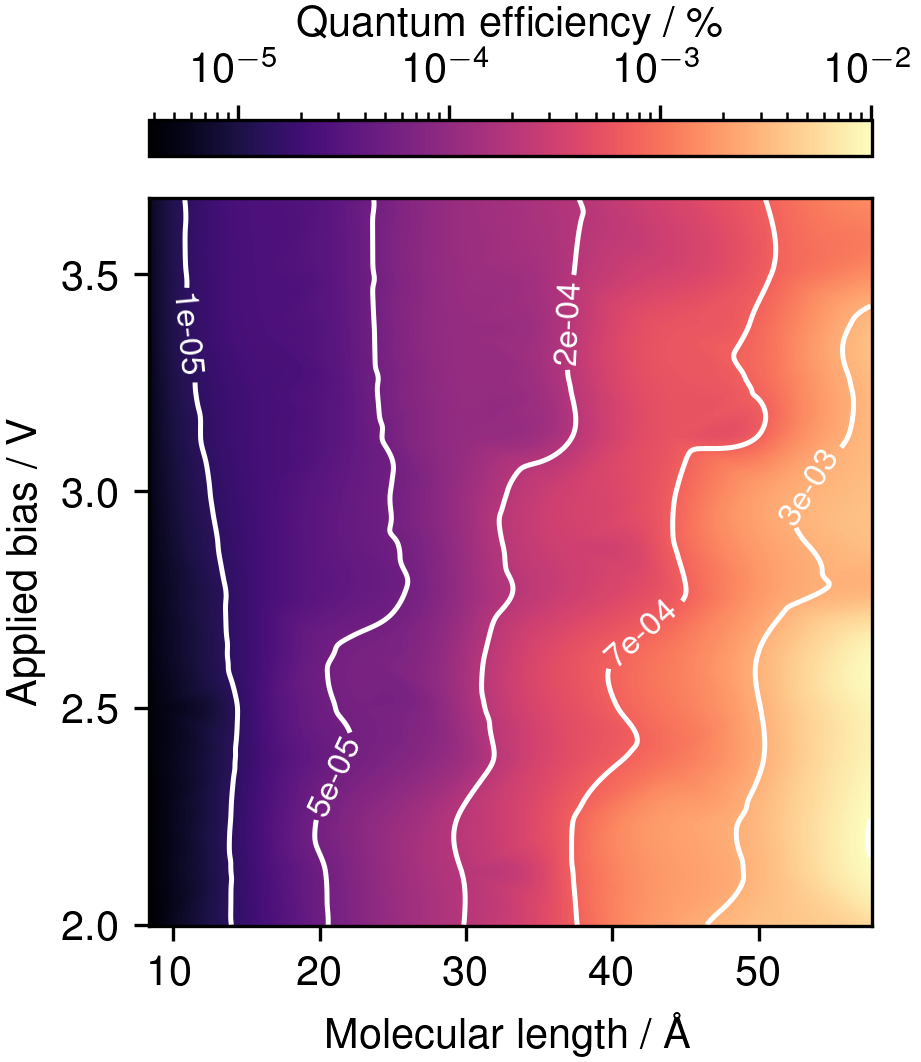}
	\caption{Color maps of the electroluminescence quantum efficiency of oligophenylenvinylene
         as a function of the applied bias
	and polymer length, for a metal-molecule coupling parameter  $\Gamma=$ 2 eV. For this coupling, long polymers and low bias maximize efficiencies.}
\label{qemap}
\end{figure}

\begin{figure*}
\centering
\includegraphics{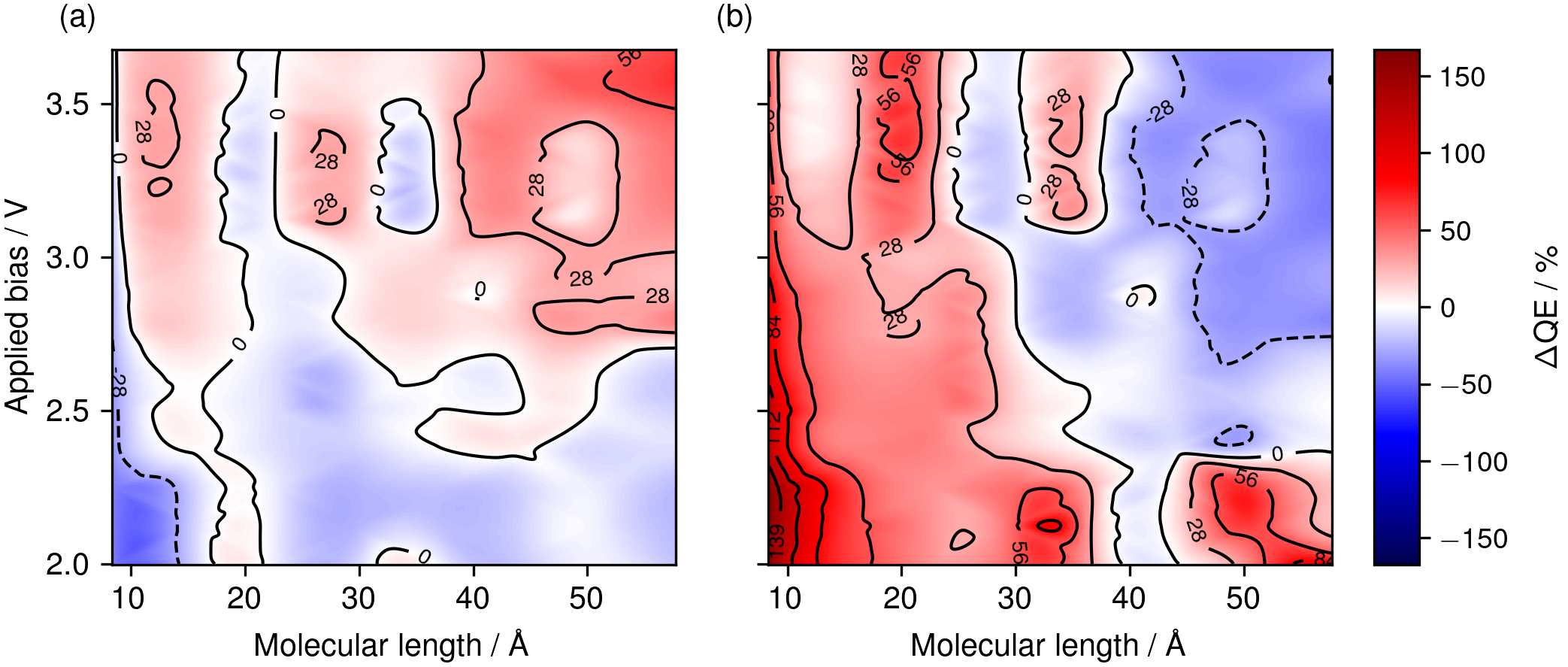}
	\caption{Effect of the coupling on the efficiency. 
	(a) Color map of the quantum efficiency for a coupling of 2.5 eV, represented as a percentage difference with
	respect to the result for $\Gamma=$ 2 eV. (b) Idem (a), but for a coupling of 1.5 eV.}
\label{qediff}
\end{figure*}

The electroluminescence quantum yields  observed for OPV derivatives 
in thin films are often higher than those found in single-molecule junctions.
Different studies have reported external quantum efficiencies between 1\% and 5\% \cite{acr_1999, macromol1999, joc_75_2599}.
These higher values in films can largely be explained by the
presence of longer polymers. In fact, from direct extrapolation of the curves plotted in Figure \ref{QvsL},
an efficiency  above 1\% can be predicted for polymers of 7 nm or longer.
Another factor contributing to  the efficiency in macroscopic assemblies is
the decoupling between the polymers and the contacts, as well as among the polymers themselves, which
enhances the ratio between emitted photons and injected carriers
by increasing resistance, as discussed above.

The color map in Figure \ref{qemap} summarizes the behavior of the quantum efficiency with the potential
and the molecular length. As a general conclusion, it is seen that the yield is maximized
with longer structures operating at low biases. Polymer length is the main factor
determining the performance, while the bias has a relatively moderate incidence.
Furthermore, weaker couplings may contribute to reach higher overall efficiencies.
This is shown in Figure \ref{qediff}, displaying 
the electroluminescent efficiencies achieved increasing
or decreasing the coupling term, relative to those obtained with $\Gamma$ = 2 eV.

\section*{Conclusion}
\label{conclusion}

This study addresses the electroluminescence of single semiconducting polymers from electron dynamics
simulations,  including an exhaustive confrontation and validation against the available experimental data. 
Our methodology proved to be suitable of correctly describing
the response of the current and the emitted power to variations in polymer length, applied bias and
coupling strength.
In particular, it reproduces the order of magnitude of key experimental parameters of 
oligophenylenvinylene in STM junctions,
such as the current, the tunneling decay constant, and the quantum efficiency of electroluminescence.

As a central result, our calculations establish that polymer length is the dominant factor in determining
the quantum efficiency, defined as the ratio between emitted photons
and injected electrons. Specifically, we find that in the tunneling regime it
increases exponentially with the length. This is a consequence of the fact that,
while the tunneling distance causes the exponential
drop of the current, its effect on the radiated power is only moderate.

The bias has a relatively small impact on the quantum yield, with an effect ruled by the power-law between emission
and current. When the emitted power scales faster than the current, increasing the bias enhances the efficiency,
and viceversa. Thus, the bias favors efficiency in short polymers but deteriorates it in long ones.
Aside from the molecular length and electronic structure, the exponent depends
on the nature of the contacts, and
therefore the coupling may determine whether the electroluminescence
efficiency is maximized at high or low applied voltages. 

The influence of the coupling strength on electroluminescence performance is not straightforward.
In principle, weakly coupled molecules can provide a higher efficiency through the reduction of the current.
On the other hand, a lower coupling decreases the tunneling constant, a behavior that has
been reported in single-molecule conductance experiments testing different  anchoring groups.
Thus, our simulations highlight a behavior that was already implied by the experimental trends,
but that has not
been explicitly recognized so far: stronger electrode-molecule couplings may yield smaller currents
in long polymers. This counterintuitive effect can be understood considering that the broadening
of the hybrid orbitals originating from a stronger metal-molecule interaction can shift the energy
of the conducting states. 

A question arises regarding the validity range for the exponential behavior of the quantum efficiency with
respect to the polymer length: this relationship must meet an upper limit, because the efficiency
can not exceed 100\%. Identifying this limit is crucial for optimizing the design of single-molecule
electroluminescent devices. According to our simulations, it must be at least 6 nm.
Where this limit lies, as well as the transition between tunneling and hopping in molecular junctions, are
important questions that challenge current methodologies and constitute the subject of ongoing
work in our group.

Overall, longer polymers operating at low bias optimize the electroluminescence quantum yield.
However, low bias implies scarce emission, leading to a compromise  between
improving efficiency and producing enough light for a given application.
Decoupling may be a resource to increase efficiency by reducing the current, but in the
case of long polymers its effect becomes less evident, since it may rise the current and also
revert the role of the bias.
By disentangling the individual contributions of the different variables determining the
quantum efficiency of electroluminescence, the present analysis
provides essential insights for complementing, interpreting, and developing
current-driven light emitting devices at the molecular level.

{\bf Acknowledgments}

This work has been funded by the European Union's Horizon
2020 research and innovation program through the project
ATLANTIC under Grant Agreement No. 823897. We also acknowledge support from CONICET (PIP 11220200103117CO) and the University of Buenos Aires (UBACyT 20020220100192BA).

\section*{Supporting Information}

\subsection*{1 Detailed methodology}

In this study, we employ a master equation, previously derived and extensively discussed by our group in Refs. \cite{jcp_fluor, jcp_electrl}. This Markovian master equation of the Redfield type models a many-electron system linearly coupled to a photon thermal bath:

\begin{equation} \label{redfield}
    \mathrm{i}\hbar \frac{\mathrm{d}}{\mathrm{d}t}\hat{\rho} = \left[\hat{H}, \hat{\rho} \right] + e \left[\hat{x}, \left[\hat{\chi}^A, \hat{\rho} \right]\right] + e\left[\hat{x}, \left( \left\{\hat{\chi}^B, \hat{\rho}\right\} + 4\hat{\rho} \text{ Tr}\left(\hat{\rho} \hat{\chi}^B \right) - 2 \hat{\rho} \hat{\chi}^B \hat{\rho} \right) \right].
\end{equation}
Here, $\hat{\rho}$ is the spinless one-electron reduced density matrix of the many-electron system, $\hat{H}$ is the electronic Hamiltonian, while $e$ denotes the elementary charge, and $\hat{x}$ is the electronic position operator. The operators $\hat{\chi}^A$ and $\hat{\chi}^B$ describe the system-bath coupling. Their matrix elements in the eigenbasis of $\hat{H}$ are given by:
\begin{align}
    \chi^A_{ij} &= -\frac{\mathrm{i}ex_{ij}|\omega_{ij}|^3}{12\pi\varepsilon_0 c^3} \left(2N\left(|\omega_{ij}| \right) + 1 \right), \\
    \chi^B_{ij} &= \frac{\mathrm{i}ex_{ij}\omega_{ij}^3}{12\pi\varepsilon_0 c^3},
\end{align}
where $\omega_{ij}= E_i - E_j$, with $E_i, E_j$ the Hamiltonian eigenvalues, $\hat{H} \ket{i} = E_i\ket{i}$, $\varepsilon_0$ is the vacuum permittivity, $c$ is the speed of light, and $N(|\omega_{ij}|)=(e^{\hbar|\omega_{ij}|/k_BT} -1)^{-1}$ is the Bose-Einstein distribution function, with $k_B$ being the Boltzmann constant and $T$ the bath temperature. A detailed microscopic derivation of Eq. \ref{redfield}, including the operators $\hat{\chi}^A$ and $\hat{\chi}^B$, can be found in the aforementioned Refs \cite{jcp_fluor, jcp_electrl}. 

The total emitted power $P$ is computed as the negative time derivative of the system's energy expectation value:
\begin{equation}
    P = -\frac{\mathrm{d}}{\mathrm{d}t}\langle \hat{H} \rangle = -2 \frac{\mathrm{d}}{\mathrm{d}t} \text{Tr} \left(\hat{\rho} \hat{H} \right),
\end{equation}
where the factor of 2 accounts for spin degeneracy. By grouping the dissipative terms of Eq. \ref{redfield} into a new operator $\hat{\Lambda}$ defined as:
\begin{equation}
    \hat{\Lambda} = e \left[\hat{x}, \left[\hat{\chi}^A, \hat{\rho} \right]\right] + e\left[\hat{x}, \left( \left\{\hat{\chi}^B, \hat{\rho}\right\} + 4\hat{\rho} \text{ Tr}\left(\hat{\rho} \hat{\chi}^B \right) - 2 \hat{\rho} \hat{\chi}^B \hat{\rho} \right) \right],
\end{equation}
the power can be expressed succinctly as:
\begin{equation} \label{power}
    P = -\frac{2}{\mathrm{i}\hbar} \text{Tr} \left(\hat{\Lambda} \hat{H} \right).
\end{equation}

Since here we are interested in the power emitted specifically from the molecular wire, we project the energy operator onto the molecular region, adopting the same strategy as in Ref. ~\cite{jcp_electrl} ~. The power radiated by the molecular subsystem $P_M$ is calculated by replacing the full Hamiltonian $\hat{H}$ in Eq. \ref{power} with a projected Hamiltonian $\hat{H}_M$:
\begin{equation} \label{eq:powproj}
    P_M = -\frac{2}{\mathrm{i}\hbar} \text{Tr} \left(\hat{\Lambda} \hat{H}_M \right),
\end{equation}
where $\hat{H}_M$ is defined according to:
\begin{equation}
    \hat{H}_M = \frac{1}{2} \left\{ \hat{P}_M, \hat{H} \right\}.
\end{equation}
Here, $\hat{P}_M$ is the projector onto the molecular region. A brief discussion of alternative projection methods can be found in Ref. \cite{jcp_electrl}. In this work, only the power emitted by the molecule is a quantity of interest. Therefore, in the main text and subsequent sections of the Supplementary Material, we drop the subscript in $P_M$ and refer to it simply as $P$.

To simulate a non-equilibrium steady state with a circulating current, we incorporate open-boundary conditions using the driven Liouville von Neumann (DLvN) method, as implemented in our previous work. \cite{jcp_electrl, etffp, tb-tddft} This involves adding an extra driving term, $\mathfrak{D}(\hat{\rho})$, to the right-hand side of the master equation (Eq. \eqref{redfield}). This term is defined in a block-matrix form corresponding to the spatial partitioning of the system (source, molecule and drain):
\begin{equation}
    \mathfrak{D}(\hat{\rho}) = -\gamma(t) 
    \begin{pmatrix}
    \hat{\rho}_S - \hat{\rho}_S^0 & \frac{1}{2} (\hat{\rho}_{SM} - \hat{\rho}_{SM}^0) & \hat{\rho}_{SD} - \hat{\rho}_{SD}^0 \\
    \frac{1}{2} (\hat{\rho}_{MS} - \hat{\rho}_{MS}^0) & 0 & \frac{1}{2} (\hat{\rho}_{MD} - \hat{\rho}_{MD}^0) \\
     \hat{\rho}_{DS} - \hat{\rho}_{DS}^0 & \frac{1}{2} (\hat{\rho}_{DM} - \hat{\rho}_{DM}^0) & \hat{\rho}_D - \hat{\rho}_D^0
\end{pmatrix}.
\end{equation}
Here, the subscripts $S, M,$ and $D$ refer to the source, molecule, and drain regions, respectively. The matrix $\hat{\rho}^0$ is a reference density matrix, taken as the ground state density matrix of the full system, polarized by the applied bias voltage. The driving rate parameter $\gamma(t)$  was smoothly ramped as a function of time to avoid inducing spurious oscillations in the density matrix. The time dependence is given by the following piecewise function:
\begin{equation}
    \gamma(t) = \begin{cases}
        \frac{\gamma_0}{2} \left[1 - \cos\left(\frac{\pi}{t_0}t \right) \right], \;\;\; &t < t_0 \\
        \gamma_0, &t \geq t_0
    \end{cases} 
\end{equation}
where the ramp time $t_0$ was fixed at 10 fs and the final driving rate was $\gamma_0 = 5\times 10^{-3}$ a.u. for all simulations. The value of $\gamma_0$ was chosen as a compromise between two requirements: it is low enough to avoid excessive artificial driving of the system, yet high enough to overcome the finite-size level spacing of the isolated leads. This ensures efficient population transfer from the electrodes to the molecular region, and prevents spurious backscattering from the boundaries. 

All molecular systems studied here are coupled to two one-dimensional metallic electrodes of 75 atoms each, attached to the right and left ends of the molecular wire, respectively.  The electronic structure of the electrodes is modeled using a nearest-neighbor tight-binding Hamiltonian in an orthonormal basis set, with a metal-metal hopping parameter $t_{MM}=-2.0$ eV, and on-site energies set to zero. 

The molecular Hamiltonian, in contrast, is computed self-consistently and includes an additional mean-field term.\cite{dlvn, jcp_electrl}
Its matrix elements are given by:
\begin{equation} \label{scfh}
    H_{ij} = t_{ij} + \delta_{ij} \sum_k \frac{2\rho_{kk} - 1}{\sqrt{R_{ik}^2+1/U^2}},
\end{equation}
where $t_{ij}$ is the hopping parameter between sites $i$ and $j$ (non-zero only for nearest neighbors). The second term is a mean-field on-site energy, representing a softened Coulomb interaction, where $R_{ik}$ is the distance between sites $i$ and $k$, and $U$ is a parameter characterizing the strength of the on-site electron-electron repulsion, fixed here at $U = 3.0$ eV.

Figure S1 shows the pattern of hopping terms used. The intramolecular hopping parameters, $t_1=-3.0$ eV and $ t_2=-2.5$ eV, were chosen to reproduce the equilibrium experimental band gap of poly(p-phenylene vinylene) (2.4 eV). \cite{Eckhardt1989, Moses2001} Of particular importance is the electrode-molecule coupling term, $\Gamma$, also indicated in Figure S1, which governs the strength of the bond between the electrode and the anchoring group and thus the electronic coupling at the interface. 

\begin{center}

\includegraphics[scale=0.5]{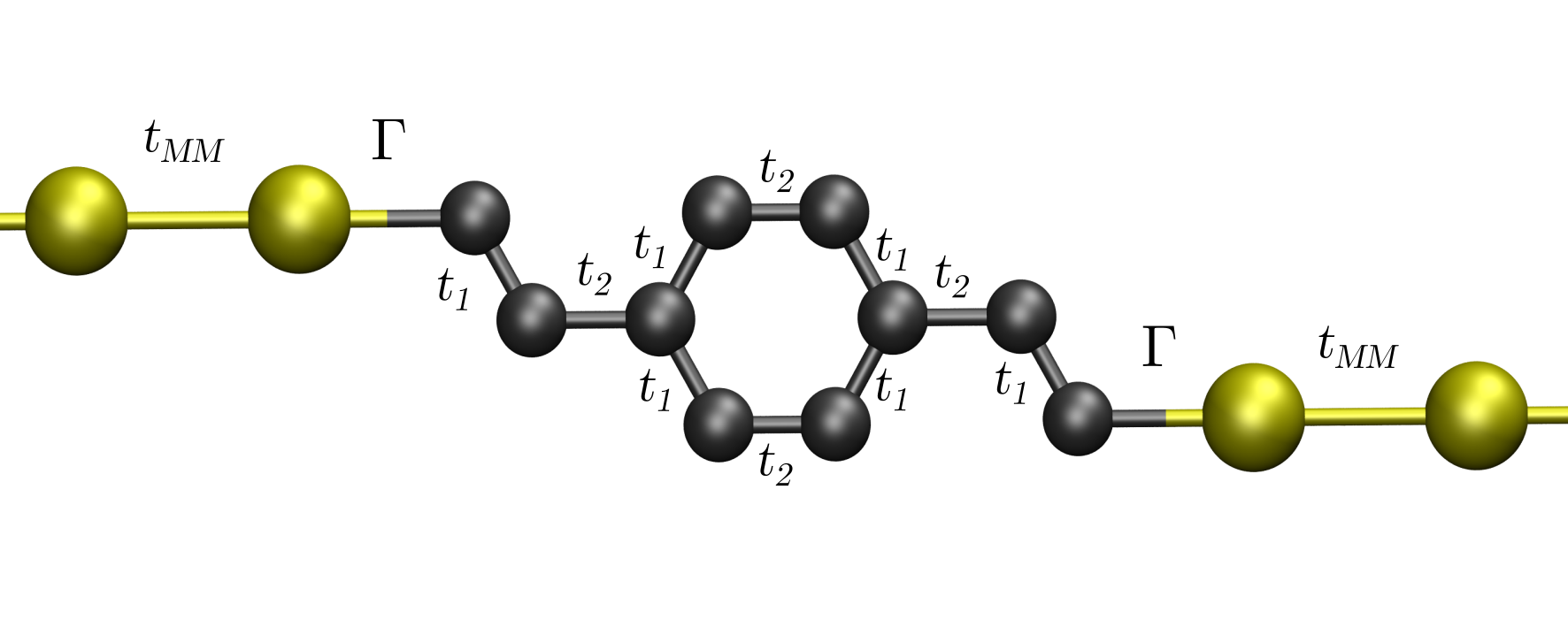}\\
	{\small FIG. S1. Schematic of the nearest-neighbor hopping terms for the shortest OPV oligomer. $t_{MM}$ denotes the metal-metal hopping, $t_1$ and $t_2$ represent the intramolecular hopping parameters, and $\Gamma$ indicates the electrode-molecule coupling strength, which is systematically varied in this study.}
\end{center}

From the calculation of the binding energy of the molecule to the electrodes, it is possible to characterize the chemical nature of the link. As found in the main manuscript, the values for this parameter ranged from $-1.5$ eV to $-2.5$ eV,
corresponding to binding energies between $\sim$25 and 70 kJ/mol. In this way the various couplings examined in the simulations mimic from electrostatic to  covalent interactions with the metal, which can be identified for example with anchoring groups as carboxylic acid in one end to thiols or amines in the other.\cite{jacs_128_15874}
We note that in the main manuscript the values for $\Gamma$ are given unsigned.

The steady-state circulating current $I$ was computed as the expectation value of the bond current operator:
\begin{equation}
    I = \text{Tr} \left( \hat{\rho} \hat{J}_{nn'}\right),
\end{equation}
where $\hat{J}_{nn'}$ describes the electron flow from site $n'$ to site $n$:\cite{jpcm_14_3049}
\begin{equation}
    \hat{J}_{nn'} = \frac{1}{\mathrm{i}\hbar} \left( \hat{P}_n \hat{H} \hat{P}_{n'} - \hat{P}_{n'} \hat{H} \hat{P}_n \right).
\end{equation}
Here, $\hat{P}_n$ is the projector onto site $n$. 

Finally, the quantum efficiency $QE$ was calculated as the ratio of emitted photons to circulating electrons, expressed as a percentage:
\begin{equation}
    \%QE = \frac{\text{Number of emitted photons}}{\text{Number of circulating electrons}}\times 100\% = \frac{P_M/E_g}{I / e} \times 100\%,
\end{equation}
where $e$ is the elementary charge and $E_g$ is the band gap of poly(p-phenylene vinylene). Therefore, $P_M/E_g$ represents the molecular photon emission rate (photons per unit time), while $I/e$ represents the electron current (electrons per unit time). 

\subsection*{2 Dependence of the tunneling decay constant on the applied bias and electrode-molecule coupling}

Figure S2 presents the tunneling decay constant, $\beta$, as a function of the applied bias voltage for a series of OPV oligomers with varying electrode-molecule coupling strengths, $\Gamma$. These $\beta$ values were obtained by computing the steady-state current for oligomers of different lengths at a constant bias and fitting the results to the expression $I = I_0 \exp(-\beta L)$, where $I_0$ is a prefactor and $L$ is the oligomer length.

\begin{center}
	\includegraphics[scale=0.75]{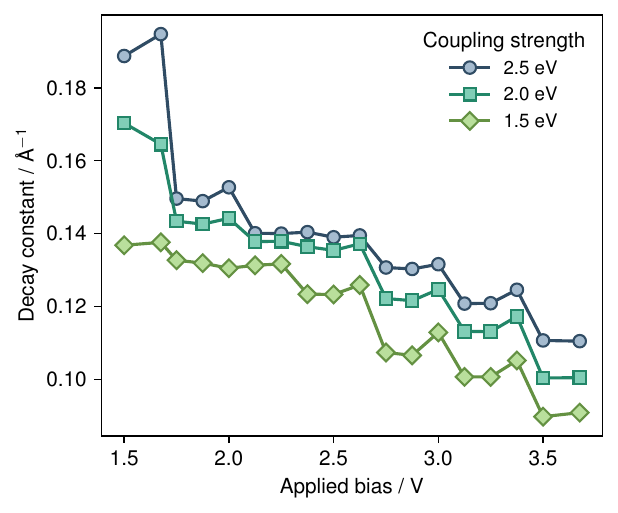}\\
	{\small FIG. S2. Tunneling decay constant $\beta$ as a function of applied bias voltage for different electrode-molecule coupling strengths ($\Gamma$). }
\end{center}

As shown in Figure S2, the tunneling decay constant generally decreases with increasing applied bias. This behavior, observed to varying extents experimentally, \cite{jacs_140_12877, naturenanot_7_713, science_301_1221, Im2022} can be attributed to two primary factors. First, the applied voltage tilts the tunneling barrier, analogous to the mechanism described by the Fowler-Nordheim model. This tilting reduces both the effective height and width of the barrier, thereby facilitating charge transport. 

Second, the decay constant decreases in a stepwise manner. These steps coincide with biases where sharp increases in current are observed (see Figure 2 in the main text), indicating that they result from new electronic states entering the bias window. Interestingly, the charge transport mechanism remains off-resonant throughout, as no regions with anomalously low decay constants are present. We explain this by field-induced localization of the molecular orbitals. Specifically, the electric field from the electrodes induces a gradient of on-site energies across the molecular wire. This promotes the formation of electronic states that are localized closer to one electrode, which enhances their coupling to that lead and consequently suppresses resonant tunneling conditions.

A second key trend is the systematic decrease in the tunneling decay constant with weaker electrode-molecule coupling. A strong contact dependence of $\beta$ has been previously predicted theoretically as a finite-size effect in molecular wires, where the nature of the chemical link can alter the current-length profile. \cite{Quek2009} While, in principle, $\beta$ could either increase or decrease with coupling strength, both experimental \cite{pccp_16_23529, Park2007,jacs_128_15874} and theoretical \cite{jpcc_113_20967} studies have consistently shown that the decay constant generally increases  with a stronger electrode-anchoring group bond, as confirmed by our data in Figure S2. We attribute this trend to the interplay of level broadening and energy-level renormalization. Stronger coupling to the electrodes leads to greater hybridization, which broadens the molecular density of states. Concurrently, this hybridization renormalizes the molecular energy levels, shifting them further from the Fermi energy. This shift increases the effective tunneling barrier, thereby resulting in a larger decay constant $\beta$ for more strongly coupled systems.

\subsection*{3 Dependence of power on molecular length and applied bias}

To qualitatively understand how the emitted power depends on various system parameters, we can express it as a sum over electronic transitions between molecular states, weighted by the population of the initial state $\rho_{ii}$:
\begin{equation} \label{eq:molpow}
    P \propto \sum_{i>j} |x_{ij}|^2 \omega_{ij}^4 \rho_{ii},
\end{equation}
where $x_{ij}$ is the transition dipole moment between states $i$ and $j$, and $\hbar\omega_{ij}=E_i-E_j$ is the transition energy. It is important to note that Eq. \ref{eq:molpow} is formally incompatible with the projected power calculation (Eq. \ref{eq:powproj}) described in Section 1, since it assumes a clear distinction between molecular and electrode states—a distinction that becomes blurred upon coupling. However, since our simulations indicate that the emitting states remain highly localized on the molecular wire, this simplified picture provides valuable physical insight.

Equation \ref{eq:molpow} indicates that the emitted power increases with larger transition dipole moments, higher transition energies, greater population in excited states, and a higher density of molecular states (which adds more terms to the summation). Figure S3 shows the steady-state emitted power as a function of molecular length for OPV oligomers under different applied bias voltages.
\begin{center}
	\includegraphics[scale=0.8]{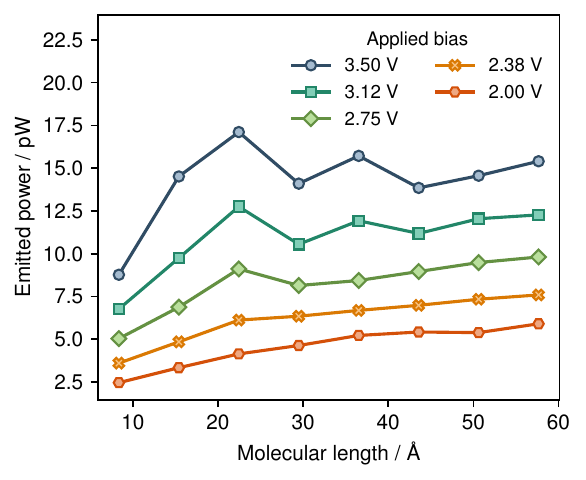}\\
	{\small FIG. S3. Emitted power as a function of molecular length at different applied bias voltages for OPV oligomers with an electrode-molecule coupling strength of 2.0 eV. }
\end{center}

At the two lowest bias voltages, the emitted power increases monotonically with molecular length. Although the steady-state current--and consequently the excited state populations $\rho_{ii}$--decreases with length at fixed bias, this reduction is outweighed by two factors: first, longer oligomers possess a higher density of electronic states, increasing the number of possible radiative transitions. Second, the enhanced conjugation in longer chains leads to more delocalized orbitals, which typically exhibit larger transition dipole moments.

This monotonic trend is disrupted at higher bias voltages (above approximately 2.5 V), where maxima appear at intermediate molecular lengths. In these cases, the enhanced power emission does not stem from anomalously high excited state populations—the transport remains in the off-resonant tunneling regime—but rather from the emergence of transitions with unusually large dipole moments. These intense transitions occur at specific molecular lengths and are bias-dependent, as they rely on electric-field-induced localization. For instance, in the trimer (22.4 Å), the bias-induced gradient of on-site energies promotes the formation of localized molecular orbitals whose transitions exhibit anomalously high intensity.

Predicting the emergence of these high-intensity transitions a priori is challenging due to the mean-field nature of the Hamiltonian and the non-equilibrium character of the steady state. Nevertheless, we observe that the prominence of these maxima diminishes in longer molecular wires.

Figure S4 shows the emitted power as a function of applied bias voltage for OPV oligomers of varying lengths. For a given oligomer, the emitted power increases with applied bias, which is expected given that a higher current is associated with a larger population of excited states, $\rho_{ii}$. Notably, the emitted power increases in a stepwise manner with bias, mirroring the behavior of the current, with steps occurring at the same voltages where the current also sharply increases. The crossing of some traces is a manifestation of the maxima at intermediate lengths discussed previously.

\begin{center}
	\includegraphics[scale=0.8]{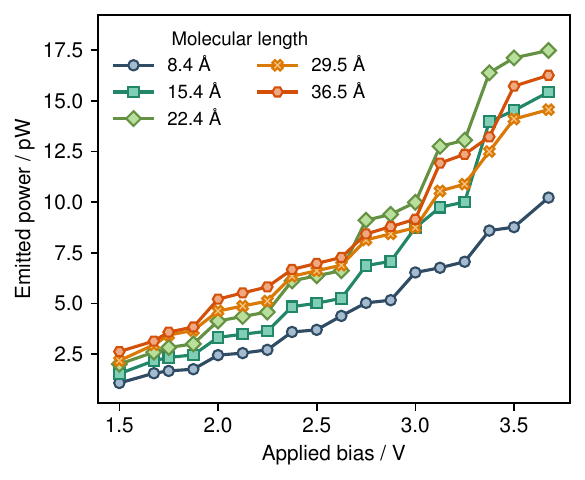}\\
{\small FIG. S4. Emitted power as a function of applied bias for OPV oligomers of different lengths with an electrode-molecule coupling strength of 2.0 eV. }
\end{center}

\subsection*{4 Dependence of the $P$-$I$ power law on electrode-molecule coupling}
As discussed in the main text, the relationship between steady-state emitted power and circulating current follows a power law of the form $P \propto I^n$, where the exponent $n$ is system-dependent. We find that $n$ depends significantly on both the molecular length and electrode-molecule coupling strength. Figures S5(a) and (b) show the steady-state power versus current for OPV oligomers of one, two and three repeating units, with electrode-molecule coupling strengths of 2.5 and 1.5 eV, respectively.

\begin{figure*}
	\centering
\includegraphics{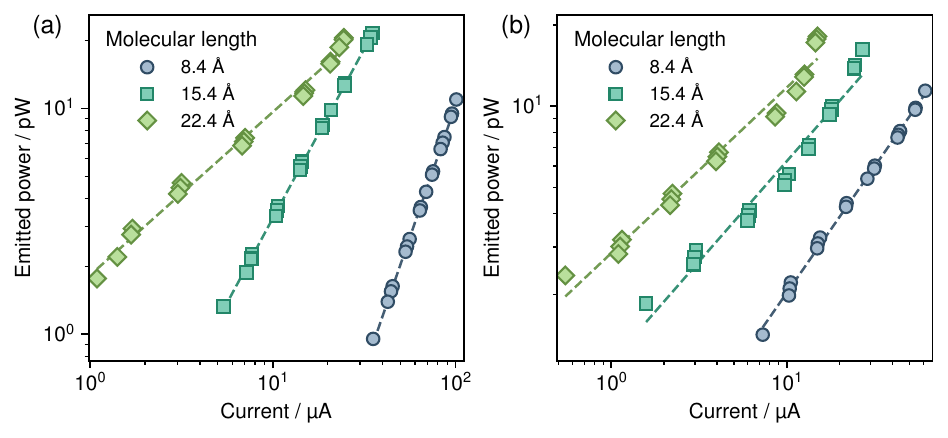}\\
{\small FIG. S5. Log-log plot of the emitted power as a function of the circulating current in the steady state. The data were fitted to a power law of the form $P \propto I^n$ to determine the exponent $n$. (a) For an electrode-molecule coupling of 2.5 eV, the fitted exponents are 2.3, 1.5 and 0.72 for increasing molecular length. (b) For a coupling of 1.5 eV, the corresponding exponents are 0.94, 0.74, and 0.61. }
\end{figure*}

\begin{table}[h]
\caption{Exponents $n$ obtained from fitting the power law $P\propto I^n$ for the first three OPV oligomers at different electrode-molecule coupling strengths $\Gamma$. }
\label{Tab:expo}
\centering
\begin{tabular}{c|ccc}
 & \multicolumn{3}{c}{$\Gamma$ / eV} \\
\cline{2-4}
\multicolumn{1}{c|}{Length / \AA} & 2.5 & 2.0 & 1.5 \\
\hline
8.4  & 2.3   & 1.5   & 0.94 \\
15.4 & 1.5   & 1.0   & 0.74 \\
22.4 & 0.72  & 0.65  & 0.61 \\
\end{tabular}
\end{table}

The fitted exponents for the first three OPV oligomers are summarized in Table \ref{Tab:expo}. The value of $n$ is crucial for understanding the bias dependence of the quantum efficiency. Since $QE \propto P/I \propto I^{n-1}$, and the system operates in a regime of positive differential conductance ($\mathrm{d}I/\mathrm{d}V >0$), an exponent $n > 1$ implies that the QE increases with bias, while $n < 1$ leads to a decrease in QE with increasing bias.

Our results demonstrate that stronger electrode-molecule couplings promote larger power-law exponents, particularly in shorter molecules. This suggests that short molecular systems can achieve higher quantum efficiencies at high operating biases when connected via strongly bonding anchoring groups. Conversely, weak couplings result in $n < 1$ even for the shortest chain, precluding any bias-induced enhancement of the QE.

Finally, we note that the influence of the electrode-molecule coupling is strongly length-dependent. For oligomers longer than 22.4 \AA, the effect of $\Gamma$ becomes less pronounced, with the power-law exponent showing a more gradual decrease and stabilizing between 0.4 and 0.6.

\bibliography{refs}

\end{document}